\documentclass[aps,twocolumn,superscriptaddress,showpacs,preprintnumbers,amsmath,amssymb]{revtex4}
\usepackage{graphicx,color,dcolumn,booktabs,bm}
\usepackage{longtable,lscape}
\usepackage{amsmath}
\usepackage{indentfirst}
\usepackage{epsfig}
\usepackage{feynmf}   
\usepackage{epstopdf}   
\usepackage{slashed}  
\usepackage{cases}
\usepackage{color}
\usepackage{multirow}
\usepackage{graphicx,color,dcolumn,booktabs,bm}
\usepackage{verbatim}
\usepackage[colorlinks,linkcolor=red,anchorcolor=green,citecolor=blue]{hyperref}
\usepackage{mathrsfs}
\usepackage{rotating}
\usepackage{threeparttable}

\newcommand{\yfos}{\Upsilon(4S)}
\newcommand{\yfis}{\Upsilon(5S)}

\newcommand{\BR}{{\cal B}}

\newcommand{\ks}{K_S^0}

\newcommand{\EE}{e^+e^-}

\newcommand{\bbc}{B^+ B^-}
\newcommand{\bbn}{B^0\bar{B}^0}

\newcommand{\infb}{\rm fb^{-1}}

\newcommand{\gev}{\rm GeV}

\newcommand{\beq}{\begin{equation}}
\newcommand{\eeq}{\end{equation}}
\newcommand{\bitm}{\begin{itemize}}
\newcommand{\eitm}{\end{itemize}}

\begin{document}

\title{\quad\\[1.0cm] Evidence for a vector charmonium-like state in \boldmath{$e^+e^- \to D^+_sD^*_{s2}(2573)^-+c.c.$}}

\noaffiliation
\affiliation{University of the Basque Country UPV/EHU, 48080 Bilbao}
\affiliation{Beihang University, Beijing 100191}
\affiliation{University of Bonn, 53115 Bonn}
\affiliation{Brookhaven National Laboratory, Upton, New York 11973}
\affiliation{Budker Institute of Nuclear Physics SB RAS, Novosibirsk 630090}
\affiliation{Faculty of Mathematics and Physics, Charles University, 121 16 Prague}
\affiliation{Chonnam National University, Gwangju 61186}
\affiliation{University of Cincinnati, Cincinnati, Ohio 45221}
\affiliation{Deutsches Elektronen--Synchrotron, 22607 Hamburg}
\affiliation{Key Laboratory of Nuclear Physics and Ion-beam Application (MOE) and Institute of Modern Physics, Fudan University, Shanghai 200443}
\affiliation{Justus-Liebig-Universit\"at Gie\ss{}en, 35392 Gie\ss{}en}
\affiliation{Gifu University, Gifu 501-1193}
\affiliation{SOKENDAI (The Graduate University for Advanced Studies), Hayama 240-0193}
\affiliation{Gyeongsang National University, Jinju 52828}
\affiliation{Department of Physics and Institute of Natural Sciences, Hanyang University, Seoul 04763}
\affiliation{University of Hawaii, Honolulu, Hawaii 96822}
\affiliation{High Energy Accelerator Research Organization (KEK), Tsukuba 305-0801}
\affiliation{J-PARC Branch, KEK Theory Center, High Energy Accelerator Research Organization (KEK), Tsukuba 305-0801}
\affiliation{Higher School of Economics (HSE), Moscow 101000}
\affiliation{Forschungszentrum J\"{u}lich, 52425 J\"{u}lich}
\affiliation{IKERBASQUE, Basque Foundation for Science, 48013 Bilbao}
\affiliation{Indian Institute of Science Education and Research Mohali, SAS Nagar, 140306}
\affiliation{Indian Institute of Technology Bhubaneswar, Satya Nagar 751007}
\affiliation{Indian Institute of Technology Hyderabad, Telangana 502285}
\affiliation{Indian Institute of Technology Madras, Chennai 600036}
\affiliation{Institute of High Energy Physics, Chinese Academy of Sciences, Beijing 100049}
\affiliation{Institute of High Energy Physics, Vienna 1050}
\affiliation{Institute for High Energy Physics, Protvino 142281}
\affiliation{INFN - Sezione di Napoli, 80126 Napoli}
\affiliation{Advanced Science Research Center, Japan Atomic Energy Agency, Naka 319-1195}
\affiliation{J. Stefan Institute, 1000 Ljubljana}
\affiliation{Institut f\"ur Experimentelle Teilchenphysik, Karlsruher Institut f\"ur Technologie, 76131 Karlsruhe}
\affiliation{Kennesaw State University, Kennesaw, Georgia 30144}
\affiliation{King Abdulaziz City for Science and Technology, Riyadh 11442}
\affiliation{Department of Physics, Faculty of Science, King Abdulaziz University, Jeddah 21589}
\affiliation{Kitasato University, Sagamihara 252-0373}
\affiliation{Korea Institute of Science and Technology Information, Daejeon 34141}
\affiliation{Korea University, Seoul 02841}
\affiliation{Kyungpook National University, Daegu 41566}
\affiliation{P.N. Lebedev Physical Institute of the Russian Academy of Sciences, Moscow 119991}
\affiliation{Faculty of Mathematics and Physics, University of Ljubljana, 1000 Ljubljana}
\affiliation{Ludwig Maximilians University, 80539 Munich}
\affiliation{Luther College, Decorah, Iowa 52101}
\affiliation{Malaviya National Institute of Technology Jaipur, Jaipur 302017}
\affiliation{University of Maribor, 2000 Maribor}
\affiliation{Max-Planck-Institut f\"ur Physik, 80805 M\"unchen}
\affiliation{School of Physics, University of Melbourne, Victoria 3010}
\affiliation{University of Mississippi, University, Mississippi 38677}
\affiliation{University of Miyazaki, Miyazaki 889-2192}
\affiliation{Moscow Physical Engineering Institute, Moscow 115409}
\affiliation{Graduate School of Science, Nagoya University, Nagoya 464-8602}
\affiliation{Universit\`{a} di Napoli Federico II, 80055 Napoli}
\affiliation{Nara Women's University, Nara 630-8506}
\affiliation{National Central University, Chung-li 32054}
\affiliation{National United University, Miao Li 36003}
\affiliation{Department of Physics, National Taiwan University, Taipei 10617}
\affiliation{H. Niewodniczanski Institute of Nuclear Physics, Krakow 31-342}
\affiliation{Nippon Dental University, Niigata 951-8580}
\affiliation{Niigata University, Niigata 950-2181}
\affiliation{University of Nova Gorica, 5000 Nova Gorica}
\affiliation{Novosibirsk State University, Novosibirsk 630090}
\affiliation{Osaka City University, Osaka 558-8585}
\affiliation{Pacific Northwest National Laboratory, Richland, Washington 99352}
\affiliation{Panjab University, Chandigarh 160014}
\affiliation{Peking University, Beijing 100871}
\affiliation{University of Pittsburgh, Pittsburgh, Pennsylvania 15260}
\affiliation{Theoretical Research Division, Nishina Center, RIKEN, Saitama 351-0198}
\affiliation{University of Science and Technology of China, Hefei 230026}
\affiliation{Seoul National University, Seoul 08826}
\affiliation{Showa Pharmaceutical University, Tokyo 194-8543}
\affiliation{Soochow University, Suzhou 215006}
\affiliation{Soongsil University, Seoul 06978}
\affiliation{University of South Carolina, Columbia, South Carolina 29208}
\affiliation{Sungkyunkwan University, Suwon 16419}
\affiliation{School of Physics, University of Sydney, New South Wales 2006}
\affiliation{Department of Physics, Faculty of Science, University of Tabuk, Tabuk 71451}
\affiliation{Department of Physics, Technische Universit\"at M\"unchen, 85748 Garching}
\affiliation{School of Physics and Astronomy, Tel Aviv University, Tel Aviv 69978}
\affiliation{Toho University, Funabashi 274-8510}
\affiliation{Department of Physics, Tohoku University, Sendai 980-8578}
\affiliation{Earthquake Research Institute, University of Tokyo, Tokyo 113-0032}
\affiliation{Department of Physics, University of Tokyo, Tokyo 113-0033}
\affiliation{Tokyo Institute of Technology, Tokyo 152-8550}
\affiliation{Tokyo Metropolitan University, Tokyo 192-0397}
\affiliation{Virginia Polytechnic Institute and State University, Blacksburg, Virginia 24061}
\affiliation{Wayne State University, Detroit, Michigan 48202}
\affiliation{Yamagata University, Yamagata 990-8560}
\affiliation{Yonsei University, Seoul 03722}
  \author{S.~Jia}\affiliation{Beihang University, Beijing 100191}\affiliation{Key Laboratory of Nuclear Physics and Ion-beam Application (MOE) and Institute of Modern Physics, Fudan University, Shanghai 200443} 
  \author{C.~P.~Shen}\affiliation{Key Laboratory of Nuclear Physics and Ion-beam Application (MOE) and Institute of Modern Physics, Fudan University, Shanghai 200443} 
  \author{I.~Adachi}\affiliation{High Energy Accelerator Research Organization (KEK), Tsukuba 305-0801}\affiliation{SOKENDAI (The Graduate University for Advanced Studies), Hayama 240-0193} 
  \author{H.~Aihara}\affiliation{Department of Physics, University of Tokyo, Tokyo 113-0033} 
  \author{S.~Al~Said}\affiliation{Department of Physics, Faculty of Science, University of Tabuk, Tabuk 71451}\affiliation{Department of Physics, Faculty of Science, King Abdulaziz University, Jeddah 21589} 
  \author{D.~M.~Asner}\affiliation{Brookhaven National Laboratory, Upton, New York 11973} 
  \author{H.~Atmacan}\affiliation{University of South Carolina, Columbia, South Carolina 29208} 
  \author{V.~Aulchenko}\affiliation{Budker Institute of Nuclear Physics SB RAS, Novosibirsk 630090}\affiliation{Novosibirsk State University, Novosibirsk 630090} 
  \author{R.~Ayad}\affiliation{Department of Physics, Faculty of Science, University of Tabuk, Tabuk 71451} 
  \author{I.~Badhrees}\affiliation{Department of Physics, Faculty of Science, University of Tabuk, Tabuk 71451}\affiliation{King Abdulaziz City for Science and Technology, Riyadh 11442} 
% ...
\collaboration{The Belle Collaboration}

\begin{abstract}
We report the measurement of $e^+e^- \to D^+_sD^*_{s2}(2573)^-+c.c.$ via initial-state radiation using a data sample of an integrated luminosity of 921.9 fb$^{-1}$ collected with the Belle detector at the $\Upsilon(4S)$ and nearby.
We find evidence for an enhancement with a 3.4$\sigma$ significance in the invariant mass of $D^+_sD^*_{s2}(2573)^- +c.c.$
The measured mass and width are $(4619.8^{+8.9}_{-8.0}({\rm stat.})\pm2.3({\rm syst.}))~{\rm MeV}/c^{2}$ and $(47.0^{+31.3}_{-14.8}({\rm stat.})\pm4.6({\rm syst.}))~{\rm MeV}$, respectively. The mass, width, and quantum numbers of this enhancement are consistent with the charmonium-like state at 4626 MeV/$c^2$ recently reported by Belle in $e^+e^-\to D^+_sD_{s1}(2536)^-+c.c.$
The product of the $e^+e^-\to D^+_sD^*_{s2}(2573)^-+c.c.$ cross section and the branching fraction of $D^*_{s2}(2573)^-\to{\bar D}^0K^-$ is measured from $D^+_sD^*_{s2}(2573)^-$ threshold to 5.6 GeV.
\end{abstract}


\maketitle

The past decade witnessed a remarkable proliferation of exotic charmonium-like and bottomonium-like resonances having properties which can not be readily explained in the framework of the expected heavy quarkonium states~\cite{C71,C74,A30,639,90,1907}.
Among the charmonium-like states, there are many vector states with
quantum numbers $J^{PC} = 1^{--}$ that are usually called $Y$
states, including the $Y(4260)$~\cite{L95,D74,L99,L110,L118}, $Y(4360)$~\cite{L98,W99,D91,D96,D89}, and $Y(4660)$~\cite{W99,D91,D96,D89,L101}.
The $Y$ states show strong coupling to hidden-charm final states, in contrast to other vector charmonium states in the same energy region, e.g., $\psi(4040)$, $\psi(4160)$, and $\psi(4415)$, which couple dominantly to open-charm meson
pairs~\cite{PDG}. These $Y$ states are good candidates for new
types of exotic particles and have stimulated many theoretical
interpretations, including tetraquarks, molecules, hybrids, and
hadrocharmonia~\cite{C71,C74,A30,639,90,1907}.

In $e^+e^-\to Y \to \pi^+\pi^-J/\psi$~\cite{L99,L110} and $\pi^+\pi^-\psi(2S)$~\cite{W99,D91} ($Y=Y(4260)$, $Y(4660)$) processes, events in the $\pi^+\pi^-$ mass spectra tend to accumulate at the nominal $f_{0}(980)$ mass, which has an $s\bar s$ component. Thus, it is natural to search for $Y$ states with a $(c{\bar s})({\bar c}s)$ quark component. Very recently, Belle reported the first vector charmonium-like state, called $Y(4626)$, decaying to a charmed-antistrange and anticharmed-strange meson pair $D^+_sD_{s1}(2536)^-+c.c.$ with a significance of 5.9$\sigma$~\cite{DsDs1}. The measured mass and width of the resonance are consistent with those of the $Y(4660)$~\cite{PDG}.
After the initial observation of the $Y(4626)$, several theoretical interpretations for this state were offered, including a molecular, diquark-antidiquark, tetraquark, or higher charmonium~\cite{ref0,ref1,ref2,ref3,ref4,ref5,ref6}.

Here, we search for $Y$ states in another charmed-antistrange and anticharmed-strange meson pair $D^+_sD^*_{s2}(2573)^-$ in $e^+e^-$  annihilations via initial-state radiation (ISR)~\cite{B27}.
The data set used in this analysis corresponds to an integrated luminosity of $921.9~\infb$ at center-of-mass (C.M.) energies of 10.52, 10.58, and 10.867 GeV collected with the Belle detector~\cite{Belle1} at the KEKB asymmetric-energy $e^+e^-$ collider~\cite{KEKB1,KEKB2}.

We use {\sc phokhara}~\cite{PHOKHARA} to generate signal Monte Carlo (MC) events. In the generator, considering that $D^+_s$ and $D^{*}_{s2}(2573)^-$ are produced from a vector state, the polar angle $\theta$ of the $D^+_s$ in the $D^+_sD^{*}_{s2}(2573)^-$ rest frame is distributed according to $(1+{\rm cos}^2\theta)$~\cite{angle1} for $e^+e^-\to D^+_sD^*_{s2}(2573)^-$, while the polar angle $\theta^{\prime}$ of the $K^-$ in the rest frame of the $D^*_{s2}(2573)^-$ is distributed according to ${\rm cos}^2\theta^{\prime}(1-{\rm cos}^2\theta^{\prime})$~\cite{angle2} for $D^*_{s2}(2573)^-\to {\bar D}^0K^-$. Generic MC samples of $\yfos \to \bbc/\bbn$, $\yfis \to B^{(*)}_s\bar{B}^{(*)}_s$, and $\EE \to q\bar{q}~(q =u,~d,~s,~c)$ at $\sqrt{s} = 10.52,~10.58$, and $10.867~\gev$ with four times the luminosity of data are used to study possible backgrounds. The detector response is simulated with GEANT3~\cite{GEANT3}.

Selections of candidates in $e^+e^- \to \gamma_{\rm ISR}D^+_s$ $D^*_{s2}(2573)^-(\to{\bar D}^0K^-)$ use well-reconstructed tracks, particle identification, and the mass-constrained fitting technique in a way similar to the methods in Ref.~\cite{Jia092015,DsDs1}. To improve the reconstruction efficiency, we fully reconstruct $\gamma_{\rm ISR}$, $D^+_s$, and $K^-$, but do not reconstruct the ${\bar D}^0$. The most energetic ISR photon is required to have energy greater than 3 GeV in the $e^+e^-$ C.M. frame. The $D^+_s$ candidates are reconstructed using the following decay modes: $\phi\pi^+$, $K^0_SK^+$, ${\bar K}^*(892)^0(\to K^-\pi^+/K^0_S\pi^0)K^+$, $\phi\rho^+$, $K^*(892)^+{\bar K}^*(892)^0(\to K^-\pi^+)$, $K^*(892)^+K^0_S$, $K^0_SK^+\pi^+\pi^-$, $\eta\pi^+$, and $\eta^{\prime}\pi^+$. Here, we select the intermediate resonances instead of the direct final states in the $D^+_s$ reconstructions in order to improve the signal-to-background ratios. The invariant masses of the $\phi(\to K^+K^-)$, $K^0_S$, $\pi^0(\to \gamma\gamma)$, ${\bar K}^*(892)^0$, $\rho^+(\to\pi^+\pi^0)$, $K^*(892)^+(\to K^+\pi^0)$, $\eta(\to \gamma\gamma)$, $\eta(\to \pi^+\pi^-\pi^0)$, and $\eta^{\prime}(\to \pi^+\pi^-\eta)$ candidates are required to be within 10, 10, 12, 50, 100, 50, 20, 10, and 10 MeV/$c^2$ of the corresponding nominal masses~\cite{PDG} ($>$90\% signal events are retained), respectively.

Next, we constrain the recoil mass of the $\gamma_{\rm ISR}D^+_sK^-$ to be the nominal mass of the ${\bar D}^0$ meson~\cite{PDG} to improve the resolution of the ISR photon energy for events within the ${\bar D}^0$ signal region (see below). As a result, the exclusive $e^+e^- \to D^+_sD^*_{s2}(2573)^-$ cross section can be measured according to the invariant mass spectrum of the $D^+_sD^*_{s2}(2573)^-$, which is equivalent to the mass of mesons recoiling against $\gamma_{\rm ISR}$.

Before calculation of the $D_s^+$ candidate mass, a fit to a common vertex is performed for charged tracks in the $D_s^+$ candidate. After the application of the above requirements,  $D^+_s$ signals are clearly observed. We define the $D^+_s$ signal region as $|M(D^+_s)-m_{D^+_s}|<12$~MeV/$c^2$ ($\sim$2$\sigma$). Here and throughout the text, $m_{i}$ represents the nominal mass of particle~$i$~\cite{PDG}.
To improve the momentum resolution of the $D^+_s$ meson candidate, a mass-constrained fit to the nominal $D^+_s$ mass~\cite{PDG} is performed. The $D^+_s$ mass sideband regions are defined as $1912.34<M(D^+_s)<1936.34$ MeV/$c^{2}$ and $2000.34<M(D^+_s)<2024.34$ MeV/$c^{2}$, each of which is twice as wide as the signal region.
The $D^+_s$ candidates from the sidebands are also constrained to the central mass values in the defined $D^+_s$ sideband regions. The $D^+_s$ candidate with the smallest $\chi^{2}$ from the $D^+_s$ mass fit is kept. Besides the selected ISR photon and $D^+_s$, we require at least one additional $K^-$ candidate in the event and retain all the combinations (the fraction of events with multiple candidates is 4\%).

\begin{figure}[htbp]
\includegraphics[width=4cm,angle=-90]{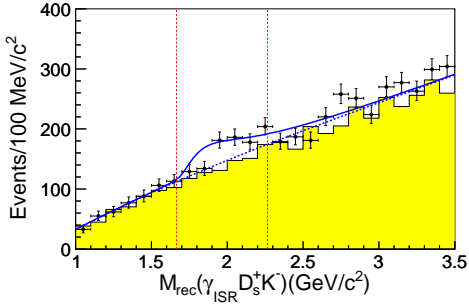}
\caption{The recoil mass spectrum against the $\gamma_{\rm ISR}D^+_sK^-$ system before applying the ${\bar D}^0$ mass constraint. The yellow histogram shows the normalized $D^*_{s2}(2573)^-$ mass sidebands (see text). The blue solid curve is the best fit, and the blue dashed curve is the fitted background. The red dashed lines show the required ${\bar D}^0$ signal region.}\label{D0}
\end{figure}

Figure~\ref{D0} shows the recoil mass spectrum against the $\gamma_{\rm ISR}D^+_sK^-$ system after requiring the events be within the $D^*_{s2}(2573)^-$ signal region (see below) in data, where the yellow histogram shows the normalized $D^*_{s2}(2573)^-$ mass sidebands (see below). The ${\bar D}^0$ signal is wide and asymmetric due to the asymmetric resolution function of the ISR photon energy and higher-order ISR corrections.
We perform a simultaneous likelihood fit to the $M_{\rm rec}(\gamma_{\rm ISR}D^+_sK^-)$ distributions of all selected $D^*_{s2}(2573)^-$ signal candidates and the normalized $D^*_{s2}(2573)^-$ mass sidebands.
The ${\bar D}^0$ signal component is modeled using a Gaussian function
convolved with a Novosibirsk function~\cite{Nov} derived from the signal MC samples, while normalized $D^*_{s2}(2573)^-$ mass sidebands are described by a second-order polynomial. The solid curve is the total fit; the ${\bar D}^0$ signal yield is $224\pm42$. An asymmetric requirement of $-200$ $<$ $M_{\rm rec}(\gamma_{\rm ISR}D^+_sK^-)-m_{{\bar D}^0}$ $<$ 400 MeV/$c^{2}$ is defined for the ${\bar D}^0$ signal region. Hereinafter the mass constraint to the recoil mass of the $\gamma_{\rm ISR}D^+_sK^-$ system is applied for events in the ${\bar D}^0$ signal region to improve the resolution of the mass.

\begin{figure}[htbp]
\includegraphics[width=4cm,angle=-90]{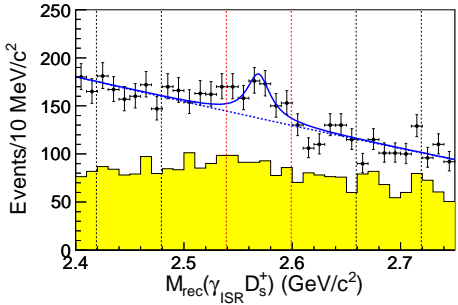}
\caption{The recoil mass spectrum
against the $\gamma_{\rm ISR}D^+_s$ system in
data. The yellow histogram shows the normalized $D^+_s$ mass
sidebands. The blue solid curve is the best fit, and the blue dashed curve is the fitted background. The red dashed lines show the required $D^*_{s2}(2573)^-$ signal region,
and the black dashed lines show the $D^*_{s2}(2573)^-$ mass sidebands.}\label{Ds2}
\end{figure}

The recoil mass spectrum against the $\gamma_{\rm ISR}D^+_s$ system after requiring the events within ${\bar D}^0$ signal region is shown in Fig.~\ref{Ds2}. A $D^*_{s2}(2573)^-$ signal is evident. The signal shape is described by a Breit-Wigner (BW) function convolved with a Gaussian function (all the parameters are fixed to those from a fit to the MC simulated distribution), and a second-order polynomial is used for the backgrounds. The fit yields $182\pm47$ $D^*_{s2}(2573)^-$ signal events as shown
in Fig.~\ref{Ds2}. We define the
$D^*_{s2}(2573)^-$ signal region as $|M_{\rm rec}(\gamma_{\rm
ISR}D^+_s)-m_{D^*_{s2}(2573)^-}|<30$~MeV/$c^2$ ($\sim$2$\sigma$),
and sideband regions as shown by blue dashed lines, each of which is twice as wide as the signal region. To
estimate the signal significance of the $D^*_{s2}(2573)^-$, we compute
$\sqrt{-2\ln(\mathcal{L}_0/\mathcal{L}_{\rm max})}$~\cite{significance},
where $\mathcal{L}_0$ and $\mathcal{L}_{\rm max}$ are the maximized likelihoods without and with the $D^*_{s2}(2573)^-$ signal, respectively. The statistical significance of the $D^*_{s2}(2573)^-$
signal is $4.1\sigma$.

\begin{figure}[htbp]
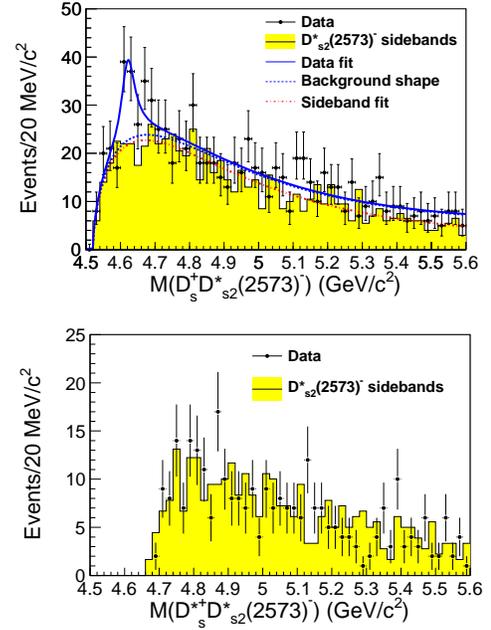

\includegraphics[width=4cm,angle=-90]{plot/fig3a.epsi}
\vspace{0.3cm}

\includegraphics[width=4cm,angle=-90]{plot/fig3b.epsi}
\caption{The $D^{+}_{s}D^*_{s2}(2573)^-$ (top) and $D^{*+}_{s}D^*_{s2}(2573)^-$ (bottom) invariant mass spectra for $e^{+}e^{-}\to D^{+}_{s}D^*_{s2}(2573)^-$ and $e^{+}e^{-}\to D^{*+}_{s}D^*_{s2}(2573)^-$.
All the components including those from the fit to the $D^{+}_{s}D^*_{s2}(2573)^-$ invariant mass spectrum
are indicated in the labels and described in the text.}\label{DsDs2}
\end{figure}

The $D^{+}_{s}D^*_{s2}(2573)^-$ invariant mass distribution is shown in Fig.~\ref{DsDs2} (top).
There is an evident peak around 4620 MeV/$c^2$, while no structure is seen in the normalized
$D^*_{s2}(2573)^-$ mass sidebands shown as the yellow histogram. In addition, no peaking background is found in the $D^+_sD^*_{s2}(2573)^-$ mass distribution from generic MC samples. Therefore, we interpret the peak in the data as evidence for a charmonium-like state decaying into $D^{+}_{s}D^*_{s2}(2573)^-$, called $Y(4620)$ hereafter.

One possible background, which is not included in the $D^*_{s2}(2573)^-$ mass sidebands,
is from $e^+e^-\to D^{*+}_{s}(\to D^+_s\gamma)D^*_{s2}(2573)^-$, where the photon from the $D^{*+}_{s}$ remains undetected.
To estimate such a background contribution, we measure this process with the data following the same procedure as used for the signal process. We require an extra photon with $E_{\gamma}>50$ MeV in the barrel or $E_{\gamma}>100$ MeV in the endcaps~\cite{ECL} to combine with the $D^+_s$ to form the $D^{*+}_s$ candidate. The mass and vertex fits are applied to the $D^{*+}_s$ candidates to improve their momentum resolutions.
In events with multiple candidates, the best candidate is chosen using the lowest $\chi^2$ value from the mass-constrained fit. The same ${\bar D}^0$ signal region requirement on $M_{\rm rec}(\gamma_{\rm ISR}D^{*+}_sK^-)$ and the ${\bar D}^0$ mass constraint are applied as in the previous analysis of $e^+e^- \to D^+_sD_{s1}(2536)^-$~\cite{Jia092015}. In the recoil mass spectrum of the $\gamma_{\rm ISR}D^{*+}_s$, $1.5\pm22.5$ $D^*_{s2}(2573)^-$ signal events are observed. After requiring the recoil mass spectrum of the $\gamma_{\rm ISR}D^+_s$ to be within the $D^*_{s2}(2573)^-$ signal region as before in $e^+e^- \to D^+_sD_{s1}(2536)^-$~\cite{Jia092015}, the $D^{*+}_{s}D^*_{s2}(2573)^-$ invariant mass distribution is shown in Fig.~\ref{DsDs2} (bottom). No evident signal is seen. The number of residual events is almost zero after subtracting the normalized $D^*_{s2}(2573)^-$  sidebands. The contribution from $e^+e^-\to D^{*+}_{s}D^*_{s2}(2573)^-$ to $e^{+}e^{-}\to D^{+}_{s}D^*_{s2}(2573)^-$ is normalized to correspond to $N^{\rm obs}_{D^{*+}_{s}D^*_{s2}(2573)^-}\varepsilon_{D^{+}_{s}D^*_{s2}(2573)^-}/\varepsilon_{D^{*+}_{s}D^*_{s2}(2573)^-}$ events. Here, $\varepsilon_{D^{+}_{s}D^*_{s2}(2573)^-}$ and $\varepsilon_{D^{*+}_{s}D^*_{s2}(2573)^-}$ are the reconstruction efficiencies of $e^+e^-\to D^{*+}_{s}D^*_{s2}(2573)^-$ to be reconstructed as $e^+e^-\to D^+_{s}D^*_{s2}(2573)^-$ and $e^+e^-\to D^{*+}_{s}D^*_{s2}(2573)^-$ to be reconstructed as $e^+e^-\to D^{*+}_{s}D^*_{s2}(2573)^-$, respectively, where the ratio of efficiencies is $(1.01\pm0.02)$, and $N^{\rm obs}_{D^{*+}_{s}D^*_{s2}(2573)^-}$ is the yield of $e^+e^-\to D^{*+}_{s}D^*_{s2}(2573)^-$ signal events in data after subtracting the normalized $D^*_{s2}(2573)^-$ sidebands and the $e^{+}e^{-}\to D^{+}_{s}D^*_{s2}(2573)^-$ background contribution. The number of normalized $e^+e^-\to D^{*+}_{s}D^*_{s2}(2573)^-$ background events in the $Y(4260)$ signal region is 1.7$\pm$1.5, which corresponds to an upper limit of 4.3 at 90\% confidence level by using the frequentist approach~\cite{Feldman} implemented in the POLE (Poissonian limit estimator) program~\cite{Conrad}.

We perform an unbinned maximum likelihood fit simultaneously to the $M(D^+_sD^*_{s2}(2573)^-)$
distributions of all selected $D^*_{s2}(2573)^-$ signal candidates and the normalized $D^*_{s2}(2573)^-$ mass
sidebands. The following components are included in the fit to the $M(D^+_sD^*_{s2}(2573)^-)$ distribution: a resonance signal, a non-resonant contribution, and the $D^*_{s2}(2573)^-$ mass sidebands. A $D$-wave BW function convolved with a Gaussian function (its width fixed at 5.0~MeV/$c^{2}$ according to the MC simulation), multiplied by an efficiency function that has a linear dependence on $M(D^{+}_{s}D^*_{s2}(2573)^-)$ and the differential ISR effective luminosity~\cite{lum} is taken as the signal shape. Here the BW formula used has the form~\cite{D95.092007}
\begin{equation} \label{eq:q1}
BW(\sqrt{s})=\frac{\sqrt{12\pi\Gamma_{ee}\BR_f\Gamma}}{s-M^2+iM\Gamma}\sqrt{\frac{\Phi_2(\sqrt{s})}{\Phi_2(M)}},
\end{equation}
where $M$ is the mass of the resonance, $\Gamma$ and $\Gamma_{ee}$ are the total width and partial width to $e^+e^-$, respectively, $\BR_f$ = $\BR(Y(4620) \to D^+_sD^*_{s2}(2573)^-)\times\BR(D^*_{s2}(2573)^-\to {\bar D}^0K^{-})$ is the product branching fraction of the $Y(4620)$ into the final state, and $\Phi_2$ is the $D$-wave two-body decay phase-space form that increases smoothly from the mass threshold with $\sqrt{s}$. The $D$-wave two-body phase space form ($\Phi_2(\sqrt{s})$) is also taken into account for the non-resonant contribution. The $D^*_{s2}(2573)^-$ mass sidebands are parameterized with a threshold function. The threshold function is
\begin{equation}\label{eq:3}
x^{\alpha}\times e^{[\beta_{1}x+\beta_{2}x^{2}]},
\end{equation}
where the parameters $\alpha$, $\beta_{1}$, and $\beta_{2}$ are free; $x$ =  $M(D^+_sD^*_{s2}(2573)^-)-x_{\rm thr}$, and the threshold parameter $x_{\rm thr}$ is fixed from generic MC simulations.

The fit results are shown in Fig.~\ref{DsDs2} (top), where the solid blue curve is the best fit,
the blue dotted curve is the sum of the backgrounds, and the red dot-dashed curve is the result of the fit to the normalized $D^*_{s2}(2573)^-$ mass sidebands. The yield of the $Y(4620)$ signal is $66^{+26}_{-20}$. The statistical significance of the $Y(4620)$ signal is $3.7\sigma$, calculated from the difference of the logarithmic likelihoods~\cite{significance}, $-2\ln(\mathcal{L}_{0}/\mathcal{L}_{\rm max}) = 19.6$, where $\mathcal{L}_{0}$ and $\mathcal{L}_{\rm max}$ are the maximized likelihoods without and with a signal component, respectively, taking into account the difference in the number of degrees of freedom ($\Delta$ndf = 3). The significance including systematic uncertainties related with the parameterization of the mass resolution, non-resonant contribution, fitted range, signal-parameterization, and efficiency function is reduced to be 3.4$\sigma$. We take this value as the signal significance. The fitted mass and width for the $Y(4620)$ are $(4619.8^{+8.9}_{-8.0}({\rm stat.})\pm2.3({\rm syst.}))~{\rm MeV}/c^{2}$ and $(47.0^{+31.3}_{-14.8}({\rm stat.})\pm4.6({\rm syst.}))~{\rm MeV}$, respectively. The value of $\Gamma_{ee}\times\BR(Y(4620) \to D^+_sD^*_{s2}(2573)^-)\times\BR(D^*_{s2}(2573)^-\to {\bar D}^0K^{-})$ is obtained to be $(14.7^{+5.9}_{-4.5}(\rm stat.)\pm3.6(syst.))$ eV. The systematic uncertainties are discussed below.

\begin{figure}[htbp]
\vspace{0.2cm}
\includegraphics[height=6.5cm,angle=-90]{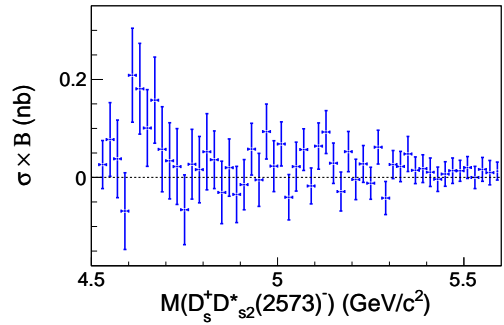}
\caption{The product of the $e^+e^-\to D^+_sD^{*}_{s2}(2573)^-$ cross section and branching fraction
$\BR(D^{*}_{s2}(2573)^-\to {\bar D}^0K^-)$ as a function of $M(D^+_sD^{*}_{s2}(2573)^-)$ with statistical uncertainties only.}\label{CS}
\end{figure}

The $e^+e^-\to D^+_sD^*_{s2}(2573)^-$ cross section is extracted from the background-subtracted $D^+_sD^*_{s2}(2573)^-$ mass distribution.
The product of the $e^+e^-\to D^+_sD^*_{s2}(2573)^-$ dressed cross section ($\sigma$)~\cite{dressed} and the decay branching fraction $\BR(D^*_{s2}(2573)^-\to {\bar D}^0K^-)$ for each $D^+_sD^*_{s2}(2573)^-$ mass bin
from threshold to 5.6 GeV/$c^2$ in steps of 20 MeV/$c^2$ is computed as
\begin{equation}\label{eq:2}
\frac{N^{\rm obs}}{\Sigma_{i}(\varepsilon_{i} \times \BR_{i}) \times \Delta{\cal{L}}},
\end{equation}
where $N^{\rm obs}$ is the number of observed $e^+e^-\to D^+_sD^*_{s2}(2573)^-$ signal events after subtracting the normalized $D^*_{s2}(2573)^-$ mass sidebands in data, $\Sigma_{i}(\varepsilon_{i} \times \BR_{i})$ is
the sum of the product of reconstruction efficiency and branching
fraction for each $D^+_s$ decay mode ($i$), and $\Delta{\cal{L}}$ is
effective luminosity in each $D^+_sD^{*}_{s2}(2573)^-$ mass bin,
respectively. The values used to calculate $\sigma(e^+e^-\to D^+_sD^{*}_{s2}(2573)^{-})\times\BR(D^{*}_{s2}(2573)^{-}\to {\bar D}^0K^{-})$ are summarized in the supplemental material~\cite{SM}. The resulting $\sigma(e^{+}e^{-}\to D^{+}_{s}D^{*}_{s2}(2573)^-) \times \BR(D^{*}_{s2}(2573)^-\to {\bar D}^{*0}K^{-})$ distribution is shown in Fig.~\ref{CS} with statistical uncertainties only.

The sources of systematic uncertainties for the cross section measurement
include detection-efficiency-related uncertainties, branching fractions of the intermediate states, the MC event generator, background subtraction, and MC statistics as well as the integrated luminosity.
The detection-efficiency-related uncertainties include those for tracking efficiency
(0.35\%/track), particle identification efficiency (1.1\%/kaon and 0.9\%/pion), $\ks$ selection efficiency
(1.4\%), $\pi^0$ reconstruction efficiency (2.25\%/$\pi^0$), and photon reconstruction
efficiency (2.0\%/photon). The above individual uncertainties from different $D^+_s$ decay channels are added linearly, and weighted by the product of the detection efficiency and $D^+_s$ branching fraction. These uncertainties are summed in quadrature to obtain the final uncertainty related to the reconstruction efficiency.
For $e^+e^-\to D^+_sD^{*}_{s2}(2573)^-$, the uncertainty from the $\theta$ dependence assumption is estimated to be 2.0\% by comparing the difference in detection efficiency between a phase space distribution and the angular distribution of $(1+{\rm cos}^2\theta)$.
Uncertainties for the $D^{+}_s$ decay branching fractions are taken from Ref.~\cite{PDG}; the final uncertainties on the $D^{+}_s$ branching fractions are summed in quadrature over all the $D^{+}_s$ decay modes weighted by the product of the efficiency and the $D^+_s$ branching fraction. The {\sc phokhara} generator calculates the ISR-photon radiator function with 0.1\% accuracy~\cite{PHOKHARA}. The uncertainty attributed to the generator can be neglected.

The systematic uncertainty associated with the combinatorial
background subtraction is due to an uncertainty in the scaling
factor (1.7\%) for the $D^{*}_{s2}(2573)^-$ sideband estimation. We
evaluate its effect on the signal yield for each bin and
conservatively assign a maximum value, 3\%.
The statistical uncertainty in the determination of efficiency from signal MC sample is about 2.0\%. The total luminosity is determined to 1.4\% uncertainty using wide-angle Bhabha scattering events. All the uncertainties are summarized in Table~\ref{systematic}. Assuming all the sources are independent, we sum them in quadrature to obtain the total systematic uncertainty.

\linespread{1.2}
\begin{table}[htbp]
\caption{Summary of the systematic uncertainties ($\sigma_{\rm syst.}$)
on the product of $e^+e^-\to D^+_sD^{*}_{s2}(2573)^-$ cross section and the decay branching fraction
$\BR(D^{*}_{s2}(2573)^-\to {\bar D}^0K^-)$.}
\vspace{0.2cm}
\label{systematic}
\begin{tabular}{c  c }
\hline
~~Source~~ & ~~~$\sigma_{\rm syst.}$~~~ \\\hline
Detection efficiency & 4.6\% \\
Branching fractions & 9.0\% \\
Background subtraction & 3.0\% \\
MC statistics  & 2.0\% \\
Luminosity & 1.4\% \\\hline
Quadratic sum & 10.9\% \\\hline
\end{tabular}
\end{table}

The following systematic uncertainties on the measured mass and width of the $Y(4620)$, and the $\Gamma_{ee}\times\BR(Y(4620) \to D^+_sD^{*}_{s2}(2573)^-)\times\BR(D^{*}_{s2}(2573)^-\to {\bar D}^0K^{-})$ are considered. The MC simulation is known to reproduce the resolution of mass peaks within 10\% over a large number of different systems. The resultant systematic uncertainties attributed to the mass resolution in the width and $\Gamma_{ee}\times\BR(Y(4620) \to D^+_sD^{*}_{s2}(2573)^-)\times\BR(D^{*}_{s2}(2573)^-\to {\bar D}^0K^{-})$ are 0.2 MeV and 0.1 eV, respectively. By changing the non-resonant background shape from a $D$-wave two-body phase space form to a threshold function, the differences of 0.2 MeV/$c^2$ and 1.9 MeV in the measured mass and width, and 0.7 eV for the $\Gamma_{ee}\times\BR(Y(4620) \to D^+_sD^{*}_{s2}(2573)^-)\times\BR(D^{*}_{s2}(2573)^-\to {\bar D}^0K^{-})$, respectively, are taken as systematic uncertainties. By changing the upper bound of the fitted range from 5.6 GeV/$c^2$ to 5.0 GeV/$c^2$, the related changes on the mass, width, and $\Gamma_{ee}\times\BR(Y(4620) \to D^+_sD^{*}_{s2}(2573)^-)\times\BR(D^{*}_{s2}(2573)^-\to {\bar D}^0K^{-})$ are 2.0 MeV/$c^2$, 3.3 MeV, and 2.3 eV. The signal-parameterization systematic uncertainty is estimated by replacing the constant total width with a mass-dependent width of $\Gamma_t$ = $\Gamma_t^0\frac{\Phi_2(M(D^+_sD^*_{s2}(2573)^-))}{\Phi_2(M_{Y(4620)})}$, where $\Gamma^0_t$ is the width of the resonance, $\Phi_2(M(D^+_sD^*_{s2}(2573)^-))$ is the phase-space form for a $D$-wave two-body system, and $\Phi_2(M_{Y(4620)})$
is the value at the $Y(4620)$ mass.
The differences in the measured $Y(4620)$ mass and width, and $\Gamma_{ee}\times\BR(Y(4620) \to D^+_sD^{*}_{s2}(2573)^-)\times\BR(D^{*}_{s2}(2573)^-\to {\bar D}^0K^{-})$ are 1.0 MeV/$c^2$, 2.3 MeV, and 2.1 eV, respectively, which are taken as the systematic uncertainties. The uncertainty in the efficiency correction from detection efficiency, branching fractions, MC statistics, and luminosity is 10.4\%. Changing the efficiency function by 10.4\% gives a 0.1 MeV/$c^2$ change on the mass, 0.2 MeV on the width, and 1.5 eV on the $\Gamma_{ee}\times\BR(Y(4620) \to D^+_sD^{*}_{s2}(2573)^-)\times\BR(D^{*}_{s2}(2573)^-\to {\bar D}^0K^{-})$. Finally, the total systematic uncertainties on the $Y(4620)$ mass, width, and $\Gamma_{ee}\times\BR(Y(4620) \to D^+_sD^{*}_{s2}(2573)^-)\times\BR(D^{*}_{s2}(2573)^-\to {\bar D}^0K^{-})$ are 2.3 MeV/$c^2$, 4.6 MeV, and 3.6 eV, respectively.

In summary, the product of the $e^+e^-\to D^+_sD^{*}_{s2}(2573)^-$ cross section and the decay branching fraction
$\BR(D^{*}_{s2}(2573)^-\to {\bar D}^0K^-)$ is measured over the C.M. energy range from the $D^+_sD^{*}_{s2}(2573)^-$ mass threshold to 5.6~GeV for the first time. We report evidence for a vector charmonium-like state decaying to $D^+_sD^{*}_{s2}(2573)^-$ with a significance of 3.4$\sigma$. The measured mass and width are $(4619.8^{+8.9}_{-8.0}({\rm stat.})\pm2.3({\rm syst.}))~{\rm MeV}/c^{2}$ and $(47.0^{+31.3}_{-14.8}({\rm stat.})\pm4.6({\rm syst.}))~{\rm MeV}$, respectively, which are consistent with the mass of $(4625.9^{+6.2}_{-6.0}({\rm stat.})\pm0.4({\rm syst.}))~{\rm MeV}/c^{2}$ and width of $(49.8^{+13.9}_{-11.5}({\rm stat.})\pm4.0({\rm syst.}))~{\rm MeV}$ of the $Y(4626)$ observed in $e^+e^-\to D^+_sD_{s1}(2536)^-$~\cite{DsDs1}, and also close to the corresponding parameters of the $Y(4660)$~\cite{PDG}.
We measure $\Gamma_{ee}\times\BR(Y(4620) \to D^+_sD^{*}_{s2}(2573)^-)\times\BR(D^{*}_{s2}(2573)^-\to {\bar D}^0K^{-})$ to be $(14.7^{+5.9}_{-4.5}(\rm stat.)\pm3.6(syst.))$ eV.

We thank the KEKB group for the excellent operation of the
accelerator; the KEK cryogenics group for the efficient
operation of the solenoid; and the KEK computer group, and the Pacific Northwest National
Laboratory (PNNL) Environmental Molecular Sciences Laboratory (EMSL)
computing group for strong computing support; and the National
Institute of Informatics, and Science Information NETwork 5 (SINET5) for
valuable network support.  We acknowledge support from
the Ministry of Education, Culture, Sports, Science, and
Technology (MEXT) of Japan, the Japan Society for the
Promotion of Science (JSPS), and the Tau-Lepton Physics
Research Center of Nagoya University;
the Australian Research Council including grants
DP180102629, 
DP170102389, 
DP170102204, 
DP150103061, 
FT130100303; 
Austrian Science Fund (FWF);
the National Natural Science Foundation of China under Contracts
No.~11435013,  
No.~11475187,  
No.~11521505,  
No.~11575017,  
No.~11675166,  
No.~11761141009;
No.~11705209;  
No.~11975076;
Key Research Program of Frontier Sciences, Chinese Academy of Sciences (CAS), Grant No.~QYZDJ-SSW-SLH011; 
the  CAS Center for Excellence in Particle Physics (CCEPP); 
the Shanghai Pujiang Program under Grant No.~18PJ1401000;  
the Ministry of Education, Youth and Sports of the Czech
Republic under Contract No.~LTT17020;
the Carl Zeiss Foundation, the Deutsche Forschungsgemeinschaft, the
Excellence Cluster Universe, and the VolkswagenStiftung;
the Department of Science and Technology of India;
the Istituto Nazionale di Fisica Nucleare of Italy;
National Research Foundation (NRF) of Korea Grant
Nos.~2016R1\-D1A1B\-01010135, 2016R1\-D1A1B\-02012900, 2018R1\-A2B\-3003643,
2018R1\-A6A1A\-06024970, 2018R1\-D1A1B\-07047294, 2019K1\-A3A7A\-09033840,
2019R1\-I1A3A\-01058933;
Radiation Science Research Institute, Foreign Large-size Research Facility Application Supporting project, the Global Science Experimental Data Hub Center of the Korea Institute of Science and Technology Information and KREONET/GLORIAD;
the Polish Ministry of Science and Higher Education and
the National Science Center; Russian Science Foundation (RSF), Grant No. 18-12-00226;
the Slovenian Research Agency;
Ikerbasque, Basque Foundation for Science, Spain;
the Swiss National Science Foundation;
the Ministry of Education and the Ministry of Science and Technology of Taiwan;
and the United States Department of Energy and the National Science Foundation.

\renewcommand{\baselinestretch}{1.2}


\begin{thebibliography}{**}
\bibitem{B27} Throughout this analysis, for any given mode, the corresponding charge-conjugate mode is implied.
\bibitem{Nov} The Novosibirsk function is defined as $f(x) =
\exp[-\frac12(\ln^2(1+\Lambda(x-x_0))/\tau^2+\tau^2)]$ with
$\Lambda=$ $\sinh(\tau\sqrt{\ln 4})/(\sigma\sqrt{\ln4})$. The
parameters represent the mean ($x_0$), the width ($\sigma$) and
the tail asymmetry ($\tau$).
\bibitem{dressed} The dressed cross section is $\sigma=\sigma^{B}/|1-\Pi|^2$, where $\sigma^{B}$ is the Born cross section, and $|1-\Pi|^2=0.94$ is the vacuum polarization factor~\cite{VP}.
\end{thebibliography}
\end{document}